\documentclass[prl,showpacs,twocolumn]{revtex4}

\usepackage{amsmath}
\usepackage{amssymb}
\usepackage[dvips]{graphicx}
\def\beq{\begin{equation}}
\def\eeq{\end{equation}}
\def\beqa{\begin{eqnarray}}
\def\eeqa{\end{eqnarray}}
\def\n{\nonumber}

\newcommand{\bel}{\begin{equation}\label}

\newcommand {\bc}{\begin{center}}
\newcommand {\ec}{\end{center}}
\newcommand {\tr}{{\rm tr}\,}

\def\dag{\dagger}

\def\vs5{\vspace*{5mm}}
\def\vs1{\vspace*{1cm}}
\def\vs2{\vspace*{2cm}}
\def\hs5{\vspace*{5mm}}
\def\hs1{\hspace*{1cm}}
\def\hs2{\hspace*{2cm}}
\def\vs50{\vspace*{50mm}}
\def\vs20{\vspace*{20mm}}
\def\tr{\hbox{tr}}

\begin{document}
\title{ Admissibility Condition and Nontrivial Indices on a Noncommutative Torus }

\author{Keiichi Nagao}
\affiliation{
Theoretical Physics Laboratory,
College of Education, Ibaraki University, 
Mito 310-8512, Japan; \\
e-mail: nagao@mx.ibaraki.ac.jp \\
}

\date{\today}


\begin{abstract}

We study the index of the Ginsparg-Wilson Dirac operator 
on a noncommutative torus numerically. 
To do this, we first formulate an admissibility condition which 
suppresses the fluctuation of gauge fields. 
Assuming this condition, we generate gauge configurations randomly, and 
find various configurations with nontrivial indices. 
We show one example of configurations with index $1$ explicitly. 
This result provides the first evidence that 
nontrivial indices can be naturally defined on the noncommutative torus 
by utilizing the Ginsparg-Wilson relation and the admissibility condition.

\end{abstract}


\maketitle

\setcounter{footnote}{0}
\paragraph{{\bf Introduction}} 
Noncommutative (NC) geometry\cite{Connes} had attracted much attention recently 
since it appears from string theory 
in $B_{\mu\nu}$ background\cite{SW}, 
and also 
from matrix models\cite{BFSS,IKKT} as their background space-time. 
Matrix models are the most promising candidates to 
formulate the superstring theory nonperturbatively. 
The construction of 
configurations with nontrivial indices in finite NC geometries or 
matrix models has been one of important subjects 
not only from mathematical interests 
but also from physical points of view. 
For instance, to realize four dimensional chiral gauge 
theories, we need to perform the Kaluza-Klein compactification of extra dimensions with 
nontrivial indices.

Topologically nontrivial configurations 
have been constructed on finite NC geometries such as fuzzy 2-sphere 
based on algebraic K-theory and projective modules in many papers, 
but the relation to indices of Dirac operators 
remains unclear in these formulations. 
We believe that the most suitable framework to discuss these problems is to utilize the 
Ginsparg-Wilson (GW) relation\cite{GinspargWilson} 
developed in lattice gauge theory.

\paragraph{{\bf General prescription}}
In ref.\cite{AIN2}, 
we proposed a general prescription to construct chirality 
and Dirac operators satisfying the GW relation and an index 
in general gauge field backgrounds on general finite NC geometries. 
The prescription proposed in ref.\cite{AIN2} is as follows. 
Let us introduce two hermitian chirality operators: 
one is a chirality operator $\gamma$, which is assumed to be independent of 
gauge fields, while the other is constructed in terms of a hermitian operator $H$ as 
\begin{equation}
\hat{\gamma} \equiv \frac{H}{\sqrt{H^2}}, \ H^{\dagger}=H.
\label{hatgamma}
\end{equation}
$\gamma$ and $\hat\gamma$ satisfy $\gamma^2=\hat{\gamma}^2=1$.  
$\hat\gamma$ depends on gauge fields through $H$. 
The Dirac operator $D_{GW}$ is defined by 
\begin{equation}
1- \gamma \hat{\gamma} = f(a,\gamma) D_{GW}, 
\label{D_GW}
\end{equation}
where $a$ is a small parameter. 
$H$ and the function $f$ must be properly chosen 
so that the $D_{GW}$ is free of species doubling and 
behaves correctly in the commutative limit ($a \rightarrow 0$). 
$D_{GW}$ satisfies the GW relation\cite{GinspargWilson}: 
\begin{equation}
\gamma D_{GW}+D_{GW} \hat{\gamma}=0.  \label{GW}
\end{equation}
Therefore the fermionic action 
$S_F={\rm tr}(\bar\Psi D_{GW} \Psi) \label{fermionic_action}$
is invariant under the modified chiral transformation~\cite{Luscher, Nieder,AIN2}
$\delta \Psi =i \lambda \hat{\gamma} \Psi, \, 
\delta \bar{\Psi} = i \bar{\Psi}\lambda \gamma$. 
The Jacobian, however, is not invariant and has the form 
\begin{equation}
q(\lambda)=
\frac{1}{2}{\cal T}r(\lambda \hat{\gamma} +\lambda \gamma), \label{top_charge}
\end{equation} 
where ${\cal T}r$ is a trace of operators acting on matrices. 
This $q(\lambda)$ is expected to provide a topological charge density, and 
the index for $\lambda=1$.

An index theorem is given by 
\begin{equation}
{\rm{index}}D_{GW}\equiv (n_+ - n_-)=\frac{1}{2} 
{\cal T}r(\gamma+\hat{\gamma}), \label{original_index}
\end{equation} 
where $n_\pm$ are numbers of zero eigenstates of $D_{GW}$ with a positive 
(or negative) chirality (for either $\gamma$ or $\hat\gamma$). 
This index theorem can be easily proven\cite{AIN3}, as done in 
lattice gauge theory\cite{Hasenfratzindex}\cite{Luscher}. 
The index defined in eq.(\ref{original_index}) is invariant 
under small deformation of any parameters such as 
gauge configurations in the operator $H$. 
We note that $\hat\gamma$ becomes singular when $H$ has zero modes. 
When an eigenvalue of $H$ crosses zero, the value of 
${\cal T}r \hat{\gamma}$ changes by two.

We may recall here that in lattice gauge theory 
the configuration space of gauge fields is topologically
trivial if we do not impose any conditions on gauge fields. 
However, noting that physically interesting gauge fields are smooth,
we can impose an admissibility 
condition\cite{Luscher:1981zq,Hernandez:1998et,Neuberger:1999pz} on gauge fields. 
This condition suppresses the fluctuation of gauge fields, and 
consequently forms a topological structure composed of isolated islands 
in the configuration space. 
This condition can also exclude zero modes of $H$\cite{Fujiwara:2000hx}. 
In ref.\cite{AIN2} we have thus expected that a similar mechanism would work 
also in finite NC geometries or matrix models, 
and that the index (\ref{original_index}) could take various integers 
according to gauge configurations. 

\paragraph{\bf The index on fuzzy 2-sphere}
In ref.\cite{AIN2} we have provided a set of simplest chirality and Dirac operators 
on fuzzy 2-sphere, as a concrete example given by the prescription.
The set in the absence of gauge fields corresponds to 
that constructed previously in ref.\cite{balaGW}. 
The nontrivial Jacobian is shown to have the correct form of 
the Chern character in the commutative limit. 
The index (\ref{original_index}), however, cannot take nonzero integers.

The authors in ref.\cite{Balachandran:2003ay} 
applied projective modules to the index (\ref{original_index}), 
so that it can take nonzero integers $\pm 1$. 
The modified index\cite{Balachandran:2003ay} 
which can take an arbitrary nonzero integer $m$\cite{AIN3} 
is symbolically expressed as 
\begin{equation}
{\text {index}} D_{GW} 
= \frac{1}{2}{\cal T}r \left\{ P^{(m)}[ A_\mu^{(m)}]
(\gamma + \hat{\gamma}[ A_\mu^{(m)}])\right\}=m. 
\label{projected_index}
\end{equation} 
The gauge fields $A_\mu^{(m)}$ are determined dependent on $m$.
$P^{(m)}$ is a projector to pick up a Hilbert space on which $\hat\gamma$ acts. 
The insertion of $P^{(m)}$ is necessary on fuzzy 2-sphere. 
The physical interpretation can be understood by considering its commutative theory. 
The configuration with $m=\pm1$\cite{Balachandran:2003ay} is interpreted as 
the 't Hooft-Polyakov monopole, and the projection expresses 
the idea of spontaneous symmetry breaking\cite{AIN3}\cite{AIMN}. 

As explained above, the index (\ref{original_index}) 
does not take nonzero integers on fuzzy 2-sphere.
Furthermore, the imposition of an admissibility condition on gauge fields, 
which can be written down so that zero modes of $H$ are excluded, 
results in providing just a vacuum sector with trivial configurations. 
This situation comes from the noncompactness of gauge fields on fuzzy 2-sphere.

In this letter we shall study the index (\ref{original_index}) 
on a NC torus, where gauge fields are defined compactly as in lattice gauge theory.
It is hence expected that 
if we formulate an admissibility condition and impose it on gauge fields, 
a topological structure would emerge in the configuration space, and 
the index (\ref{original_index}) could take nonzero integers. 
This conjecture motivates us to study the index (\ref{original_index}) on a NC torus. 

\paragraph{\bf The GW Dirac operator on a NC torus}
A $d=2n$ dimensional NC torus is formulated by using $L\times L$ dimensional 
't Hooft matrices ($L^d=N^2$)\cite{CDS,Ambjorn:1999ts}. 
To concentrate on explaining our main result, we refer the formulation of the NC torus 
to ref.\cite{Iso:2002jc}, and  follow all conventions therein.
Then let us begin with introducing the GW Dirac operator on the NC torus. 
Since the NC torus has a lattice structure\cite{Ambjorn:1999ts}, 
we can use the overlap Dirac operator\cite{Neuberger:1998fp}, 
which is a practical solution to the GW relation in lattice gauge theory, 
by replacing lattice difference operators with their matrix correspondences 
on the NC torus\cite{Nishimura:2001dq}.

The GW Dirac operator $D_{GW}$\cite{Neuberger:1998fp} 
on the $d=2n$ dimensional NC torus is given by eq.(\ref{D_GW}) with 
setting $H$, $\gamma$ and $f$ in eqs.(\ref{hatgamma})(\ref{D_GW}) as follows :
$\gamma=\gamma_{d+1}$, $f=a$, 
$H = \gamma_{d+1}(m_0-aD_{\rm w})$, 
$D_{\rm w}={1\over2}\left[\gamma_\mu(\nabla_\mu^*+\nabla_\mu)
-ar\nabla_\mu^*\nabla_\mu\right]$. 
$\gamma_{d+1}$ is an ordinary chirality operator, and 
$a$ is a lattice spacing, which is defined by the relation 
$L_{phys}=La$, where $L_{phys}$ is the size of the NC torus.
$m_0$ and~$r$ are free parameters. 
The Dirac operator in the absence of gauge fields is free of species doubling 
if $0<\frac{m_0}{r}<2$. In the following we set $r=1$ for simplicity. 
$\nabla_\mu$ and $\nabla_\mu^*$ are forward and backward covariant 
difference operators respectively. They act on matrices as follows: 
$\nabla_\mu\psi={1\over a}
\left[U_\mu \Gamma_\mu (\Gamma_\mu^\dag)^R -1 \right] \psi $,
$\nabla_\mu^*\psi={1\over a}
\left[1-\Gamma_\mu^\dag U_\mu^\dag \Gamma_\mu^R \right] \psi$ ,
where the superscript $R$ means that the operator
acts on matrices from the right.
$U_{\mu}$ are analogues of link variables 
and $\psi$ is a fermion
in the fundamental representation of the gauge group. 
$\Gamma_\mu$ is a shift operator which is constructed by taking a $d/2$ tensor 
product of 't Hooft matrices. 
The explicit representation is given in ref.\cite{Iso:2002jc}.
$U_\mu$, $\Gamma_\mu$ and $\psi$ are $N\times N$ matrices.
$D_{GW}$ transforms covariantly under the gauge transformation 
since $\psi$, $U_\mu$ and $\lambda$ transforms as follows: 
$\psi \rightarrow g \psi$,
$U_\mu \rightarrow g U_\mu \Gamma_\mu g^\dag \Gamma_\mu^\dag$,
$\nabla_\mu \psi \rightarrow g \nabla_\mu \psi$, 
$\nabla_\mu^* \psi \rightarrow g \nabla_\mu^* \psi$, 
$\lambda \rightarrow g \lambda g^\dag$.
We assume the gauge group to be abelian in the following calculation, which 
can be easily generalized to the nonabelian gauge group.

The nontrivial Jacobian (\ref{top_charge}) on the NC torus 
is shown to have the form of the Chern character 
with star-products in a weak coupling expansion\cite{Iso:2002jc} 
by utilizing a topological argument in ref.\cite{Fujiwara:2002xh}. 
A parity anomaly is derived in ref.\cite{Nishimura:2002hw}. 
The index theorem on the NC torus is given by eq.(\ref{original_index}). 
We expect that the index (\ref{original_index}) can take nonzero integers 
according to gauge configurations  
if we impose an admissibility condition 
which suppresses the fluctuation of gauge fields. 
It would be better if we could write down an analytical expression of 
gauge configurations with nontrivial indices explicitly. 
Such an expression, however, has not been discovered yet. 
Therefore we shall analyze the index numerically. 
To do this, we first formulate an admissibility condition.

\paragraph{\bf An admissibility condition on a NC torus}
The gauge action on the NC torus is given by 
\begin{equation}
S_G=N \beta \sum_{\mu > \nu} \tr 
\left[ 1-\frac{1}{2}(P_{\mu\nu} + P_{\mu\nu}^\dag) \right]. \label{gauge action}
\end{equation}
$P_{\mu\nu}$ is the plaquette which is expressed as 
\begin{equation}
P_{\mu\nu}
= Z_{\mu\nu}^* V_\mu V_\nu V_\mu^\dag V_\nu^\dag \label{plaquette} , 
\end{equation}
where we have introduced $V_\mu=U_\mu \Gamma_\mu$ and 
$Z_{\mu\nu}=\exp(-i\frac{2\pi}{L}) \epsilon_{\mu\nu}$. 
This is the twisted version\cite{TEK} of the Eguchi-Kawai model\cite{EK}, 
which was shown to be a nonperturbative description of 
NC Yang-Mills theory\cite{NCMM}\cite{Ambjorn:1999ts}. 

For an admissibility condition on the NC torus, we assume the following 
gauge-invariant expression:
\begin{equation}
\Vert 1- P_{\mu\nu} \Vert < \eta_{\mu\nu} \quad {\text {for all} } \,\, \, \mu > \nu .
\label{adm_NCtorus}
\end{equation}
The norm $\Vert \Vert$ of a matrix $O$ is defined as 
$\Vert {O} \Vert \equiv
\left[ {\text {the maximal eigenvalue of }}O^\dag O \right]^{\frac{1}{2}} $. 
$\eta_{\mu\nu}$ are some positive parameters which should be chosen appropriately. 
The condition (\ref{adm_NCtorus}) implies 
$\Vert \left[ \nabla_\mu , \nabla_\nu \right] \Vert < \eta_{\mu\nu}/a^2$. 
This is the bound on the field strength, which becomes irrelevant in the continuum 
limit. In this sense the condition (\ref{adm_NCtorus}) is physically natural. 
Applying arguments in refs.\cite{Hernandez:1998et,Neuberger:1999pz} 
onto the NC torus, 
we can show that zero modes of $H$ are excluded, namely, the minimal eigenvalue of 
$H^2$ is larger than zero, if we choose $\eta_{\mu\nu}$ such that 
$\sum_{\mu>\nu} \eta_{\mu\nu} \le \frac{1}{2+\sqrt{2}} \left\{ 1-(1-m_0)^2 \right\}$.
If we take $\eta_{\mu\nu}=\eta$, the inequality reads 
$\eta \le \frac{(2-\sqrt{2})}{d(d-1)} \left\{ 1-(1-m_0)^2 \right\} \equiv \eta_{max}$.
In the following we set $\eta_{\mu\nu}=\eta_{max}$.
It is noteworthy that the upper bound on $\Vert H \Vert$ can be shown as 
\begin{equation}
\Vert H \Vert \le |m_0-d| + d. \label{upper_bound}
\end{equation}

\paragraph{\bf Nontrivial indices on a NC torus}
We shall analyze the index (\ref{original_index}) numerically 
under the admissibility condition (\ref{adm_NCtorus}) with $\eta_{\mu\nu}=\eta_{max}$.
The index can be estimated by evaluating the eigenvalues of $H$. 
In fact the index is equal to half of the difference of 
the number of the positive eigenvalues and that of the negative ones. 
To evaluate the eigenvalues of $H$, 
we write down a matrix representation of $H$ :
\begin{equation}
H \Rightarrow
\begin{pmatrix}
H_{\vec k^{(1)},\vec k'^{(1)}} & H_{\vec k^{(1)},\vec k'^{(2)}}  
& \cdots & H_{\vec k^{(1)},\vec k'^{(L^2)}} \cr 
\cdot & H_{\vec k^{(2)},\vec k'^{(2)}}  & \cdots 
&H_{\vec k^{(2)},\vec k'^{(L^2)}} \cr  
\cdot & \cdot & \cdots  & \cdots \cr
\cdot & \cdot & \cdot &H_{\vec k^{(L^2)},\vec k'^{(L^2)}}  \cr 
\end{pmatrix} ,
\end{equation}
where the lower triangular components are abbreviated since $H$ is hermitian. 
Each component $H_{\vec k , \vec k'}$, where 
the superscripts of $\vec k$ and $\vec k'$ are omitted, is 
a $2^{\frac{d}{2}} \times 2^{\frac{d}{2}}$ matrix spanned by spinor indices: 
\begin{eqnarray}
&&\frac{1}{N} \langle s | \tr \left[ 
\exp(-ik_\nu \hat x_\nu) H \exp \left( i k'_\rho \hat x_\rho \right) \right] 
|s' \rangle \n \\
&=&\langle s | \gamma_{d+1} |s' \rangle \n \\
&& \otimes \frac{r}{2} 
\sum_\mu \left\{ \frac{1}{N^2} 
U_\mu(k-k') 
\exp\left(i \left[ \frac{1}{2}k'_\rho k_\sigma \theta_{\rho\sigma} 
+k'_\mu a \right] \right) 
\right. \n \\
&&+ \frac{1}{N^2} 
U_\mu^*(k'-k) \exp \left( i \left[ \frac{1}{2}k'_\rho k_\sigma \theta_{\rho\sigma} 
-k_\mu a \right] \right) \n \\
&&
\left.
+(m_0 -2)\delta_{\vec k \vec k'}
\right\} 
\quad -\sum_\mu \langle s | \gamma_{d+1} \gamma_\mu |s' \rangle \n \\
&&
\otimes \frac{1}{2N^2} 
\left\{ 
U_\mu(k-k') \exp \left( i \left[ \frac{1}{2}k'_\rho k_\sigma \theta_{\rho\sigma} 
+k'_\mu a \right] \right) 
\right. \n \\
&&-
\left.
U_\mu^*(k'-k) \exp\left( i \left[ \frac{1}{2}k'_\rho k_\sigma \theta_{\rho\sigma} 
-k_\mu a \right] \right)
\right\}. \label{comp_H}
\end{eqnarray}
In the first line $\tr$ does not act on spinor indices, and 
$e^{i \vec k \cdot \vec{\hat x}}=
\exp\left( i \frac{2\pi \vec m}{L_{phys}}\cdot \vec{\hat x} \right)$ 
is a NC plane wave. 
$\vec m$ is an integral vector in ${\bold Z}^d$ modulo $L$. 
$\vec{\hat x}$ is a formal expression of a hermitian NC coordinate satisfying 
$[\hat x_\mu,\hat x_\nu]=i \theta_{\mu\nu}$, where $\theta_{\mu\nu}$ 
is a NC parameter. We leave the detail to ref.\cite{Iso:2002jc}. 
In the left-hand side of eq.(\ref{comp_H}) 
$s$ and $s'$ represent the spinor indices, and 
$\frac{1}{N}$ is a normalization factor.

Using the matrix representation of $H$, 
we shall look for configurations with nontrivial indices 
under the admissibility condition (\ref{adm_NCtorus}). 
We generate many configurations of 
$U_\mu=\frac{1}{N^2}\sum_{\vec m \in ({\bold Z}^d )_L} 
U_\mu(k) e^{i \vec k \cdot \vec{\hat x}}$, 
which is linked to the corresponding field 
$U_\mu(x)=\frac{1}{N^2}\sum_{\vec m \in ({\bold Z}^d )_L} 
U_\mu(k) e^{i \vec k \cdot \vec x}$ 
through $U_\mu(k)$, 
by utilizing random numbers on a computer. 
Extracting configurations 
satisfying the admissibility condition (\ref{adm_NCtorus}),
we evaluate the eigenvalues of $H$ for each configuration by using LAPACK. 
We thus obtain the index for each configuration.  
Then the configurations with nontrivial indices can be picked up. 
According to this prescription, we have analyzed the index on 
the simplest $d=2$ dimensional NC torus for various $L$(=$N$) with setting $r=m_0=a=1$. 
We have thus found various configurations with nontrivial indices for $L \ge 3$.

In the case of $L=4$, for instance, we have discovered configurations 
with indices $0$,$\pm 1$,$\cdots$,$\pm 4$.
One example of configurations with index $1$ 
is exhibited in the left-center column of table \ref{tab1}. 
In the right column thirty-two eigenvalues of $H$ are listed. 
The range of the eigenvalues is contained in the closed segment $[-3,3]$, which is 
consistent with (\ref{upper_bound}).
The number of positive eigenvalues is seventeen, while that of negative ones is fifteen. 
We thus see that the index is given by $1$. 
\begin{table}
\caption{A configuration with index 1 }
\label{tab1}
{\footnotesize
\begin{tabular}{|c|c||c|} \hline
{$U_\mu (x_1,x_2)$} & (Re $U_\mu (x_1,x_2)$, Im $U_\mu (x_1,x_2)$) 
& e.v. of $H$ \\ \hline
 $U_1(0,0)$ & (0.727201133,-0.686424481) & -2.70790584 \\
 $U_1(0,1)$ & (-0.891526109,0.452969143) & -2.26895517 \\
 $U_1(0,2)$ & (-0.480794165,0.876833706) & -2.1997985 \\
 $U_1(0,3)$ & (-0.959813891,-0.280637005) & -2.06401734 \\
 $U_1(1,0)$ & (0.727201013,-0.686424608) & -1.91158358\\
 $U_1(1,1)$ & (-0.89152603,0.452969299) & -1.81010727 \\
 $U_1(1,2)$ & (-0.480794012,0.87683379) & -1.64921482\\
 $U_1(1,3)$ & (-0.95981394,-0.280636837) & -1.5717787 \\
 $U_1(2,0)$ & (0.727200893,-0.686424736) & -1.40574954 \\
 $U_1(2,1)$ & (-0.891525951,0.452969455) & -1.31040394 \\
 $U_1(2,2)$ & (-0.480793858,0.876833874) & -0.986584996 \\
 $U_1(2,3)$ & (-0.959813989,-0.280636669) & -0.817042275 \\
 $U_1(3,0)$ & (0.727200773,-0.686424863) & -0.256864337 \\
 $U_1(3,1)$ & (-0.891525872,0.452969611) & -0.239882409 \\
 $U_1(3,2)$ & (-0.480793705,0.876833958) & -0.131022138 \\
 $U_1(3,3)$ & (-0.959814038,-0.280636501) & 0.126736602 \\  
 $U_2(0,0)$ & (-0.793082477,-0.609114355) & 0.228034989 \\
 $U_2(0,1)$ & (0.997805689,-0.0662108309) & 0.269224886 \\
 $U_2(0,2)$ & (-0.899357079,-0.437214664) & 0.382605754 \\
 $U_2(0,3)$ & (-0.346180482,0.938167999) & 0.393858179 \\
 $U_2(1,0)$ & (-0.793082583,-0.609114217) & 0.618954906\\
 $U_2(1,1)$ & (0.997805677,-0.0662110054) & 0.923470992 \\
 $U_2(1,2)$ & (-0.899357156,-0.437214507) & 1.0013276\\
 $U_2(1,3)$ & (-0.346180318,0.93816806) & 1.22666957 \\
 $U_2(2,0)$ & (-0.79308269,-0.609114078) & 1.30262876 \\
 $U_2(2,1)$ & (0.997805666,-0.0662111799) & 1.45624418 \\
 $U_2(2,2)$ & (-0.899357232,-0.43721435) & 1.9123136 \\
 $U_2(2,3)$ & (-0.346180154,0.93816812) & 2.03997661 \\
 $U_2(3,0)$ & (-0.793082796,-0.609113939) & 2.1969756 \\
 $U_2(3,1)$ & (0.997805654,-0.0662113543) & 2.23883707 \\
 $U_2(3,2)$ & (-0.899357308,-0.437214193) & 2.29084575 \\
 $U_2(3,3)$ & (-0.34617999,0.938168181) & 2.72220583 \\ \hline
\end{tabular}
}
\end{table}
For our purpose here it is sufficient to find and show at least one configuration 
with a nontrivial index. 
We do not have to check whether the configuration minimizes 
the gauge action (\ref{gauge action}), 
since the index (\ref{original_index}) is topologically 
invariant against small deformation of configurations. 
Our observation has confirmed 
the existence of topologically nontrivial configurations, and 
that the index (\ref{original_index}) can take nonzero integers on the NC torus.


\paragraph{{\bf Discussions}}
To summarize, we have formulated an admissibility condition which suppresses 
the fluctuation of gauge fields on a NC torus by following the construction of it 
in lattice gauge theory. 
Next we have numerically studied the index of the GW Dirac operator 
under the admissibility condition on the simplest $d=2$ dimensional NC torus 
for various $L(=N)$. 
We have found various configurations with nontrivial indices. 
As a concrete example we have exhibited a configuration with index 1 
in the case of $L=4$. 
This is the first evidence which has confirmed that nontrivial indices can be 
naturally realized on the NC torus by utilizing the GW relation and 
the admissibility condition.

Further investigation of the index on the NC torus is necessary to 
understand the topological structure of gauge fields 
under the admissibility condition. 
In lattice gauge theory the admissibility 
condition in refs.\cite{Luscher:1981zq,Hernandez:1998et,Neuberger:1999pz} 
has been numerically analyzed\cite{Fukaya:2003ph,Fukaya:2005cw,Bietenholz:2005rd}. 
On the NC torus, the gauge action accommodated 
to the admissibility condition can be written down as 
$S_G=N \beta \sum_{\mu > \nu} \tr
\left[ \frac{1-(P_{\mu\nu} + P_{\mu\nu}^\dag)/2}
{1- \Vert 1- P_{\mu\nu} \Vert / \eta_{\mu\nu}}
\right]$ if $\Vert 1- P_{\mu\nu} \Vert < \eta_{\mu\nu}$, and $S_G=\infty$ otherwise, 
by following the construction of it in lattice gauge theory\cite{Luscher:1998du}. 
It would be interesting to perform similar studies of 
the admissibility condition on the NC torus. 
On the other hand, analytical studies are also needed. 
For example, an analytical expression of nontrivial 
configurations on the NC torus remains unknown. In the context 
of the recent study of QCD\cite{Kiskis:2002gr} based on  
the quenched version\cite{Bhanot:1982sh} of the Eguchi-Kawai model\cite{EK}, 
such an expression is derived\cite{Kikukawa:2002ms}. 
These kinds of study are indispensable to construct 
chiral gauge theories\cite{Nishimura:2001dq} on the NC torus. 
Analyses on other NC geometries such as fuzzy 2-sphere are also important and 
should be continued. 
We expect that our formalism can provide a clue for classifying 
the topology of not only gauge configuration space but also 
space-time\cite{Hanada:2005vr} 
since space-time and matter are indivisible in matrix models or NC field theories. 
We hope to report some progress in these directions in the future.

The author would like to thank H.~Aoki, H.~Fuji, S.~Iso, H.~Suzuki and S.~Watamura 
for valuable discussion, and T.~Azuma, W.~Bietenholz, Y.~Kikukawa, Y.~Kitazawa and 
J.~Nishimura for helpful comments. 
The content of this letter was presented by the author 
at the meeting of the Physical Society of Japan held in Miyazaki on Sep.9-12(10), 2003.


\begin{thebibliography}{all}

\bibitem{Connes}
A. Connes, Noncommutative geometry, Academic Press, 1990.

\bibitem{SW} N.~Seiberg and E.~Witten,
JHEP {\bf 9909}, 032 (1999).

\bibitem{BFSS}
T.~Banks, W.~Fischler, S.~H.~Shenker and L.~Susskind,
Phys.\ Rev.\ D {\bf 55}, 5112 (1997).

\bibitem{IKKT}
N.~Ishibashi, H.~Kawai, Y.~Kitazawa and A.~Tsuchiya,
Nucl.\ Phys.\ B {\bf 498}, 467 (1997). 



\bibitem{GinspargWilson}P.~H.~Ginsparg and K.~G.~Wilson,
Phys.\ Rev.\ D {\bf 25}, 2649 (1982).

%

\bibitem{AIN2}
H.~Aoki, S.~Iso and K.~Nagao,
Phys.\ Rev.\ D {\bf 67}, 085005 (2003). 
%

\bibitem{Luscher}M.~L\"uscher,
Phys.\ Lett.\ B {\bf 428}, 342 (1998).

\bibitem{Nieder}F.~Niedermayer,
Nucl.\ Phys.\ Proc.\ Suppl.\  {\bf 73}, 105 (1999).

\bibitem{AIN3}
H.~Aoki, S.~Iso and K.~Nagao,
Nucl.\ Phys.\ B {\bf 684}, 162 (2004).

\bibitem{Hasenfratzindex}P.~Hasenfratz,
Nucl.\ Phys.\ Proc.\ Suppl.\  {\bf 63}, 53 (1998);
%
P.~Hasenfratz, V.~Laliena and F.~Niedermayer,
Phys.\ Lett.\ B {\bf 427}, 125 (1998).


\bibitem{Luscher:1981zq}
M.~L\"uscher,
Commun.\ Math.\ Phys.\  {\bf 85}, 39 (1982).

\bibitem{Hernandez:1998et}
P.~Hernandez, K.~Jansen and M.~L\"uscher,
Nucl.\ Phys.\ B {\bf 552}, 363 (1999).

\bibitem{Neuberger:1999pz}
  H.~Neuberger,
  Phys.\ Rev.\ D {\bf 61}, 085015 (2000).


\bibitem{Fujiwara:2000hx}
  T.~Fujiwara,
  Prog.\ Theor.\ Phys.\  {\bf 107}, 163 (2002).


\bibitem{balaGW}
A.~P.~Balachandran, T.~R.~Govindarajan and B.~Ydri,
hep-th/0006216.



\bibitem{Balachandran:2003ay}
A.~P.~Balachandran and G.~Immirzi,
Phys.\ Rev.\ D {\bf 68}, 065023 (2003).

\bibitem{AIMN}
  H.~Aoki, S.~Iso, T.~Maeda and K.~Nagao,
  Phys.\ Rev.\ D {\bf 71}, 045017 (2005)
  [Erratum-ibid.\ D {\bf 71}, 069905 (2005)].


\bibitem{CDS} A.~Connes, M.~R.~Douglas and A.~Schwarz,
JHEP {\bf 9802}, 003 (1998).
 
\bibitem{Ambjorn:1999ts}
  J.~Ambjorn, Y.~M.~Makeenko, J.~Nishimura and R.~J.~Szabo,
  JHEP {\bf 9911}, 029 (1999);
%
  Phys.\ Lett.\ B {\bf 480}, 399 (2000);
%
  JHEP {\bf 0005}, 023 (2000).


\bibitem{Iso:2002jc}
S.~Iso and K.~Nagao,
Prog.\ Theor.\ Phys.\  {\bf 109}, 1017 (2003).


\bibitem{Neuberger:1998fp}H.~Neuberger,
Phys. Lett. B{\bf 417},141 (1998);
Phys. Lett. B{\bf427},353 (1998).



\bibitem{Nishimura:2001dq}
J.~Nishimura and M.~A.~Vazquez-Mozo,
JHEP {\bf 0108}, 033 (2001).
%

\bibitem{Fujiwara:2002xh}
T.~Fujiwara, K.~Nagao and H.~Suzuki,
JHEP {\bf 0209}, 025 (2002).


\bibitem{Nishimura:2002hw}
  J.~Nishimura and M.~A.~Vazquez-Mozo,
  JHEP {\bf 0301}, 075 (2003).



\bibitem{TEK}
A.\ Gonz\'{a}lez-Arroyo and M.\ Okawa,
Phys. Rev. D {\bf 27} (1983) 2397.


\bibitem{EK} 
T.\ Eguchi and H.\ Kawai, 
Phys.\ Rev.\ Lett. {\bf 48} (1982) 1063.


\bibitem{NCMM}
H.~Aoki, N.~Ishibashi, S.~Iso, H.~Kawai, Y.~Kitazawa and T.~Tada,
Nucl.\ Phys.\ B {\bf 565}, 176 (2000).



\bibitem{Fukaya:2003ph}
  H.~Fukaya and T.~Onogi,
  Phys.\ Rev.\ D {\bf 68}, 074503 (2003).


  
  
  
\bibitem{Fukaya:2005cw}
  H.~Fukaya, S.~Hashimoto, T.~Hirohashi, K.~Ogawa and T.~Onogi,
  Phys.\ Rev.\ D {\bf 73}, 014503 (2006).


\bibitem{Bietenholz:2005rd}
  W.~Bietenholz, K.~Jansen, K.~I.~Nagai, S.~Necco, L.~Scorzato and S.~Shcheredin,
  hep-lat/0511016.



\bibitem{Luscher:1998du}
  M.~L\"uscher,
  Nucl.\ Phys.\ B {\bf 549}, 295 (1999).


\bibitem{Kiskis:2002gr}
J.~Kiskis, R.~Narayanan and H.~Neuberger,
Phys.\ Rev.\ D {\bf 66}, 025019 (2002).

\bibitem{Bhanot:1982sh}
G.~Bhanot, U.~M.~Heller and H.~Neuberger,
Phys.\ Lett.\ B {\bf 113}, 47 (1982).

\bibitem{Kikukawa:2002ms}
Y.~Kikukawa and H.~Suzuki,
JHEP {\bf 0209}, 032 (2002).


\bibitem{Hanada:2005vr}
  M.~Hanada, H.~Kawai and Y.~Kimura,
  Prog.\ Theor.\ Phys.\  {\bf 114}, 1295 (2005).



\end{thebibliography}
\end{document}